\documentclass[a4paper]{jpconf}
\usepackage{graphicx}
\usepackage{subfigure}
\begin{document}
  \title{Jet measurements in proton-proton collisions with the ALICE experiment at LHC}
  \author{Michal Vajzer$^{1,2}$ on behalf of the ALICE collaboration}
  \address{$^1$Nuclear Physics Institute -- Prague, Czech Republic}
  \address{$^2$Faculty of Nuclear Sciences and Physical Engineering, CTU -- Prague, Czech Republic}
  \ead{michal.vajzer@cern.ch}

  \begin{abstract}
 The study of jets, collimated sprays of particles associated with hard partons, is an important tool in testing perturbative quantum chromodynamics (pQCD) and probing hot and dense nuclear matter created in high energy heavy-ion collisions. 
 Jets enable the  study of hard scatterings, fragmentation and hadronisation and their modification in the presence of a nuclear medium with respect to baseline vacuum measurements, which is acquired from jet measurements in proton-proton collisions.

 We have analysed data from proton-proton collisions at $\sqrt{s}\rm{ = 7}$~TeV measured by the ALICE experiment at the LHC and reconstructed the inclusive spectra of charged particle jets at mid-rapidity using anti-$k_{\rm{T}}$ clustering algorithm. 
 We present the jet spectra corrected for detector effects using several unfolding methods. Furthermore, we examine various properties of jets, such as their charged particle multiplicity and jet shapes.
  \end{abstract}

  \section{Introduction}

  Jets are collimated sprays of particles created from an energetic parton (quark or gluon), which is produced in the initial stage of the collision.  
They are an excellent tool to test pQCD and study fragmentation and hadronisation in proton-proton collisions and their modification in collisions of heavy ions. 
Here, they also serve as probe of hot and dense nuclear matter with an aim of exploring quark-gluon plasma (QGP), which is a challenging task due to an elusive characteristics that requires a combination of various signatures.

A Large Ion Colliding Experiment (ALICE) has excellent tracking capabilities and can reconstruct and identify charged particles with transverse momenta down to 150~MeV/$c$. 
We report measurements of the charged jet spectra and jet shape observables, such as charged track multiplicities and radial momentum density distributions~\cite{sidharth}. 
These observables are studied at centre of mass energy $\sqrt{s} \rm{= 7}$~TeV and compared to predictions obtained from Monte-Carlo generators.
 
  \section{Data analysis}

The results presented in this paper are based on an analysis of the data from proton-proton collisions at the LHC taken in 2010 by the ALICE. 
The online event selection is done using the V-ZERO (V0) detector and Inner Tracking System (ITS), selecting 160M minimum bias events. 
From these events, only those with reconstructed primary vertex within 10~cm from the nominal interaction point are further analysed.  
    
For charged jet reconstruction, tracks from the ITS and Time Projection Chamber (TPC) are used. 
These tracks are selected to be within pseudorapidity $\left| \eta \right| \le 0.9$ and transverse momenta ($p_{\rm{T}}$) greater than 0.15~GeV/$c$. 
They are used as input for FastJet~\cite{FastJet} to reconstruct jets using anti-$k_{\rm{T}}$ algorithm. 
It belongs to sequential recombination jet algorithms. 
Anti-$k_{\rm{T}}$ merges hard particles first while $k_{\rm{T}}$ starts with soft particles and is therefore more suitable for determination of background. 
Resolution parameters $R = \sqrt{\Delta \eta ^2 + \Delta \varphi ^2 }$, used in this analysis are 0.2, 0.4 and 0.6. 
For further analysis, jets had to have $p_{\rm{T}}$ of at least 5~GeV/$c$ and be within $\left| \eta_{jet} \right| \le 0.9 - R$. 

\begin{figure}[b]
	\begin{center}
	\includegraphics[width = \textwidth]{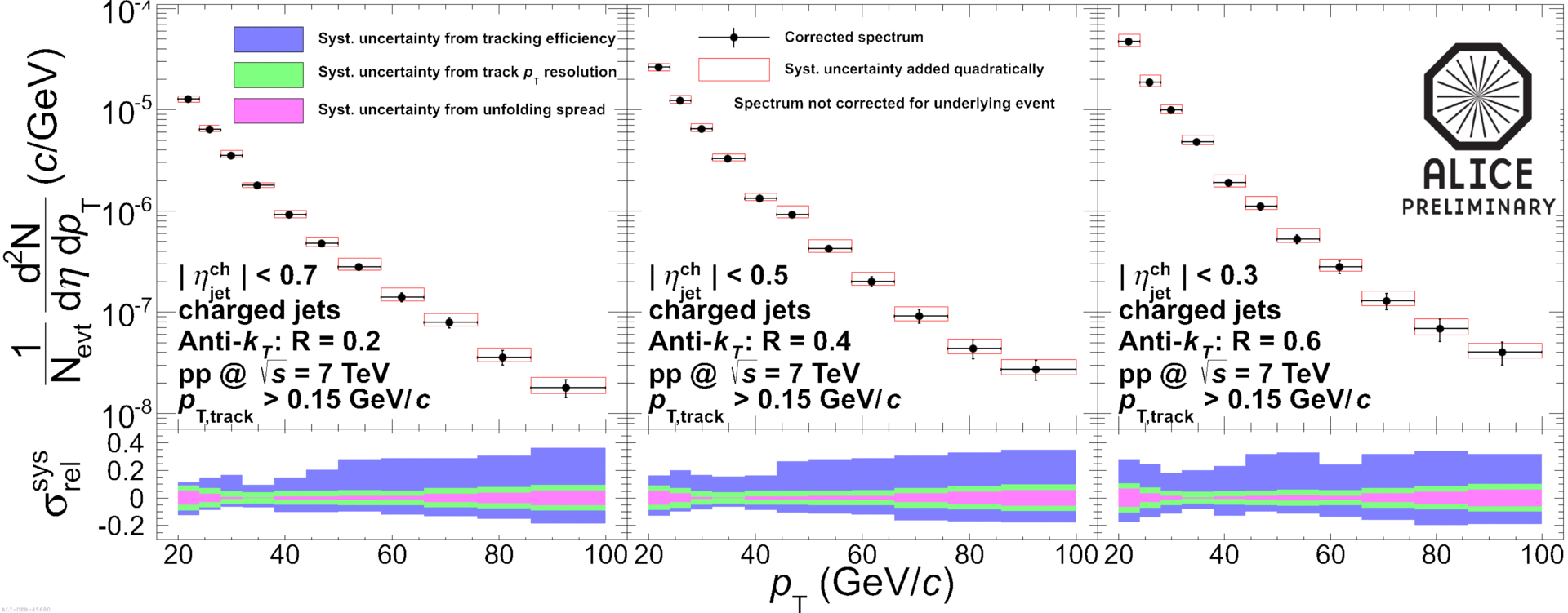}
	\caption{ (Color online) Charged jet spectra corrected for detector effects reconstructed using anti-$k_{\rm{T}}$ algorithm with resolution parameter $R$  (left) 0.2; and (middle) 0.4; and (right) 0.6; from pp collisions at $\sqrt{s} \rm{= 7}$~TeV. Error bars indicate statistical uncertainties. In bottom part are relative systematic uncertainties obtained from different sources for the anti-$k_{\rm{T}}$ algorithm with corresponding $R$. }
	\label{res-spectra}
	\end{center}
\end{figure}

In order to correct effects of detector response and to obtain charged jet spectra at particle level and corrected jet shape observables.
To do this we are using PYTHIA~\cite{pythia}~(Perugia0 tune) simulated data to obtain spectra on a particle level and a full GEANT simulation of detector to extract detector response matrices linking generator level jet spectra with recontructed spectra.
We use 3 different correction procedures. 

As PYTHIA reproduces the measured jet observables such as spectra and jet shapes, reasonably well, as seen in~\cite{sidharth}, we adopted  a bin-by-bin correction procedure using simulated events as first method. 
In this technique, the correction factors (CF) are computed for each jet $p_{\rm{T}}$ bin as the ratio between the observable at particle level and the same observable at a detector level in the same $p_{\rm{T}}$ bin, 
\begin{equation}
 \rm{CF}\left(\textit{p}\rm{_T}\right) =  \left. \frac{ \rm{d}\sigma\rm{^{ch}_{jet}} }{ \rm{d}\textit{p}\rm{{_T}d}\eta } \right|_{MC}^{part}  \Big/  \left. \frac{ \rm{d}\sigma\rm{^{ch}_{jet}} }{ \rm{d}\textit{p}\rm{{_T}d}\eta } \right|_{MC}^{det}.  
 \label{unfold-eq-bbb}
\end{equation}
This correction is subsequently applied to a raw jet spectrum. 

Second method, we used to find an inverse matrix to the response matrix, is Bayesian unfolding~\cite{unfold-bayes}. 
This is possible because the response matrix is composed of a probability distributions of finding given a cause -- generated jet of given $p\rm{^{ch,MC-gen}_{T,jet}}$ -- as an effect -- reconstructed jet with any value $p\rm{^{ch,MC-rec}_{T,jet}}$. 

Third method and another regularized unfolding procedure used as a cross-check in jet spectrum analysis is the Singular Value Decomposition of the response matrix~\cite{unfold-svd}. 
This method decomposes the response matrix into two square orthogonal matrices and a diagonal singular matrix. 
Subsequently it converts the problem to a system of linear equations and solves it.

  \section{Results}

In Fig.~\ref{res-spectra} we present the unfolded charged jet spectra for anti-$k_{\rm{T}}$ algorithm with resolution parameters $R$ of 0.2, 0.4 and 0.6 in top panel, with the relative systematic uncertainties in the bottom panel of each plot, reconstructed from proton-proton collisions at $\sqrt{s}\rm{= 7}$~TeV using the ALICE detector system. 
The most significant uncertainties are obtained from varying track reconstruction effiency, track $p\rm{_T}$ resolution and simulation with parametrised detector response. 
The final systematic uncertainties are obtained by adding above mentioned uncertainties in quadrature. 

\begin{figure}[t]
	\begin{center}
	\subfigure[Anti-$k_{\rm{T}}$; R = 0.4]{\includegraphics[width = 0.40\textwidth]{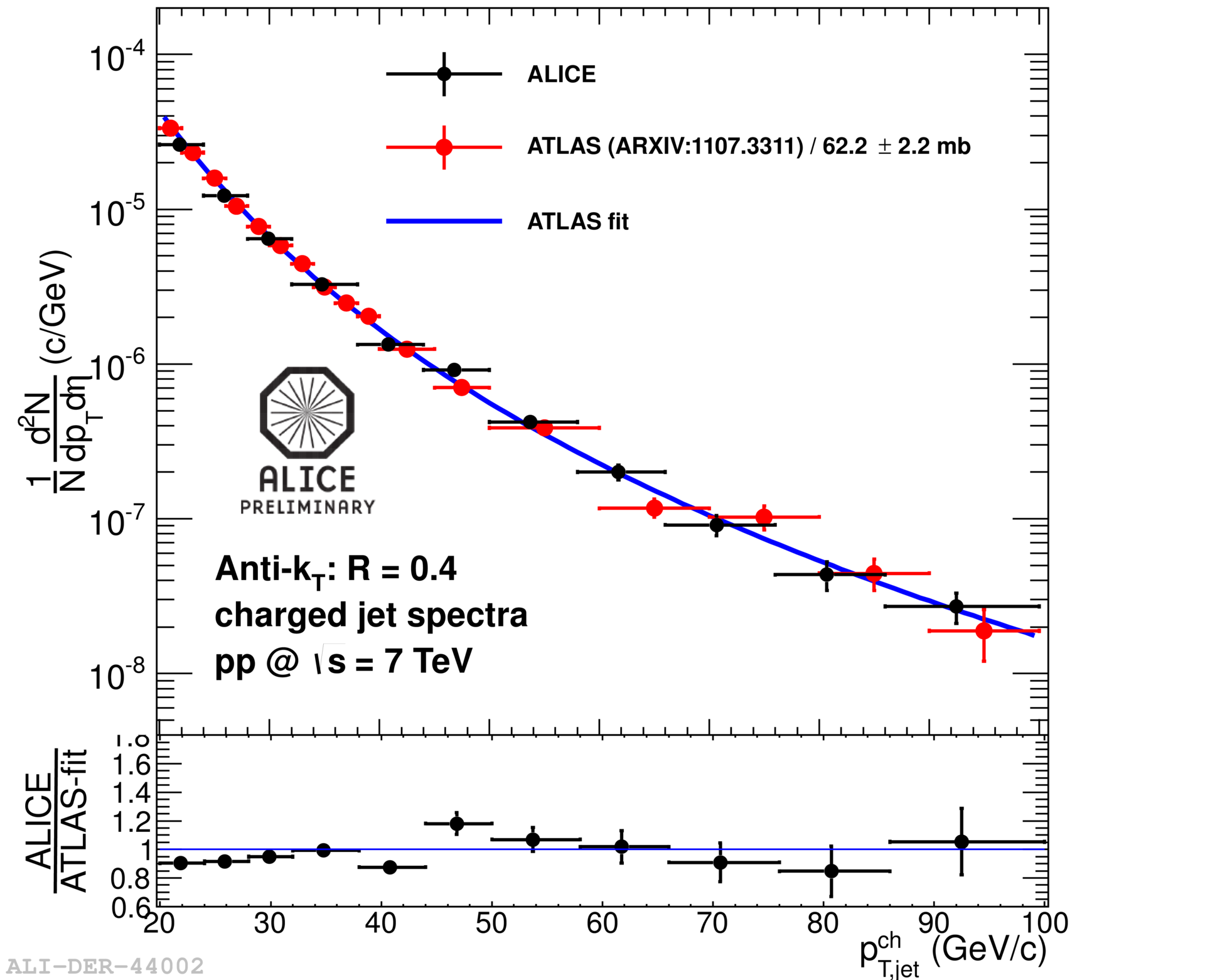}}
	\subfigure[Anti-$k_{\rm{T}}$; R = 0.6]{\includegraphics[width = 0.40\textwidth]{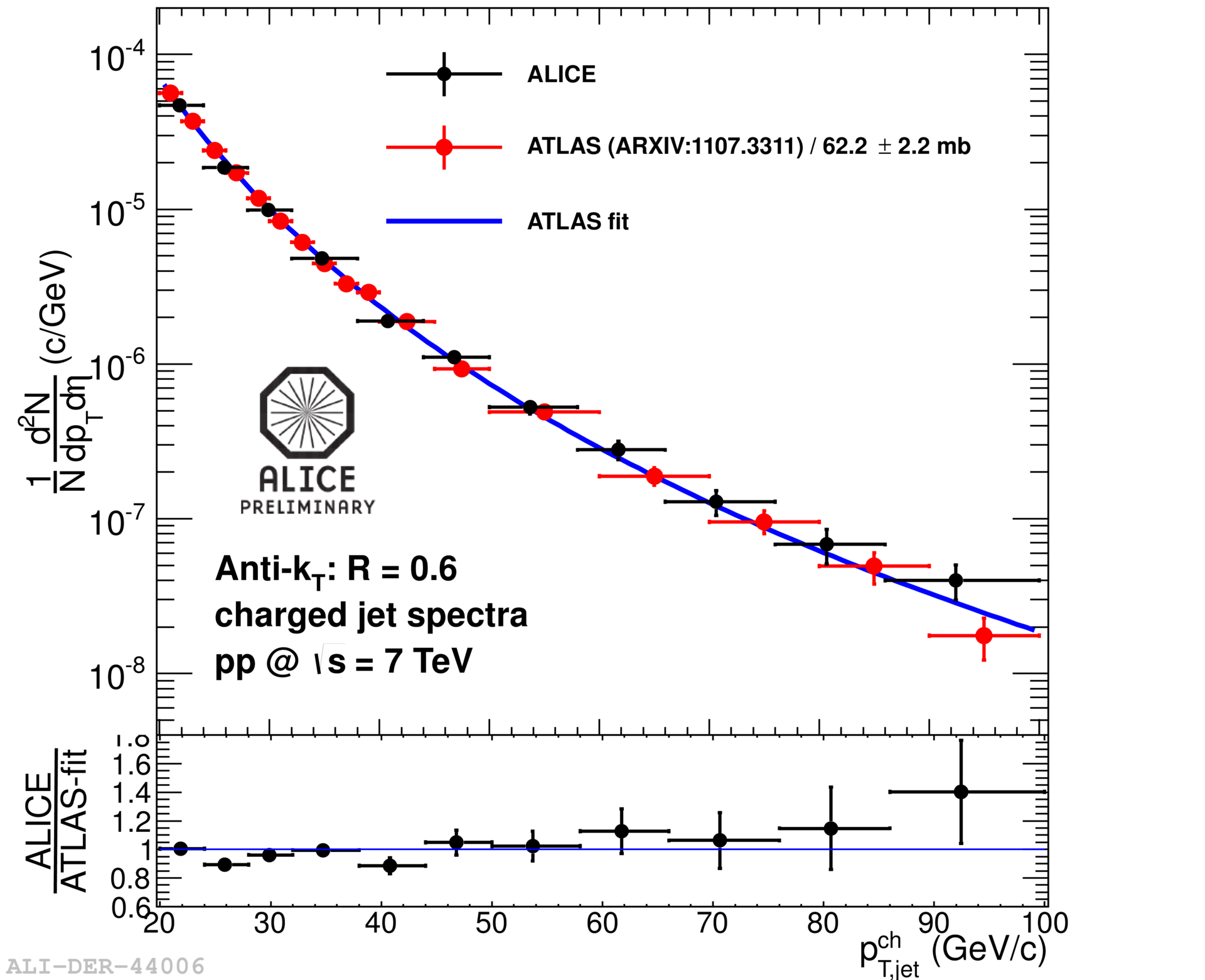}}
	\caption{ (Color online) Charged jet spectra from anti-$k_{\rm{T}}$ algorithm with resolution parameter $R$ of (a) 0.4; and (b) 0.6; from pp collisions at $\sqrt{s} \rm{= 7}$~TeV measured by the ALICE (black points) and the ATLAS (red points) collaborations. The line is a Tsallis fit of the ATLAS results. Error bars indicate statistical uncertainties. In bottom part is ratio of the ALICE charged jet spectrum to the ATLAS results. }
	\label{res-spectra-ATLAS}
	\end{center}
\end{figure}

The results presented here are consistent with results obtained by the ATLAS collaboration~\cite{ATLAS-charged-jet-spectra}, as shown in the Fig~\ref{res-spectra-ATLAS}, obtained from charged tracks in $\rm{\left| \eta
\right| \le 0.5}$ with $p\rm{_{T} \ge 0.3}$~GeV/$c$.

\begin{figure}[t]
	\begin{center}
	\subfigure[Charged track multiplicity]{\label{res-tracks} \includegraphics[width = 0.40\textwidth]{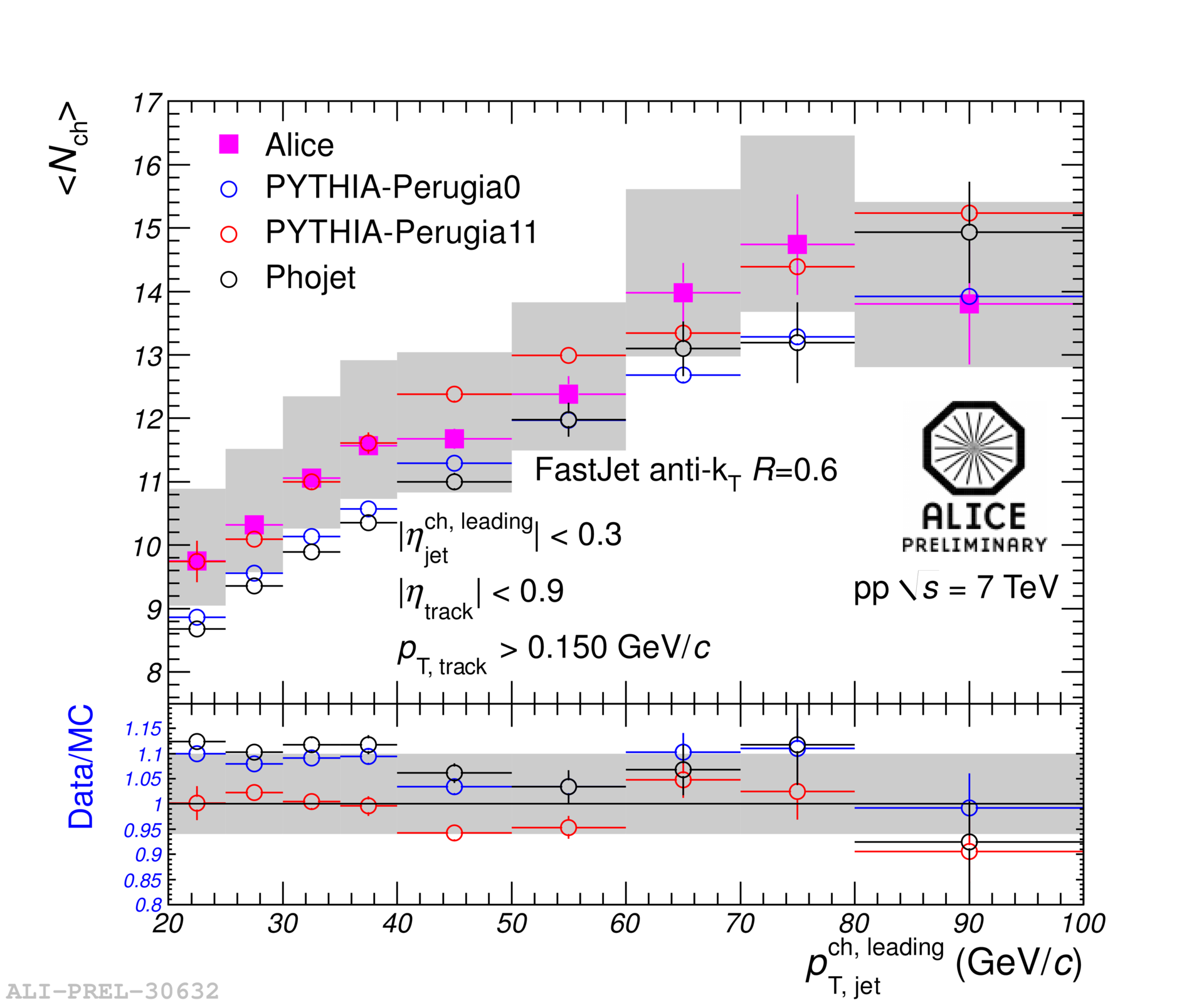}}
	\subfigure[Radial momentum distribution]{\label{res-tracks1} \includegraphics[width = 0.40\textwidth]{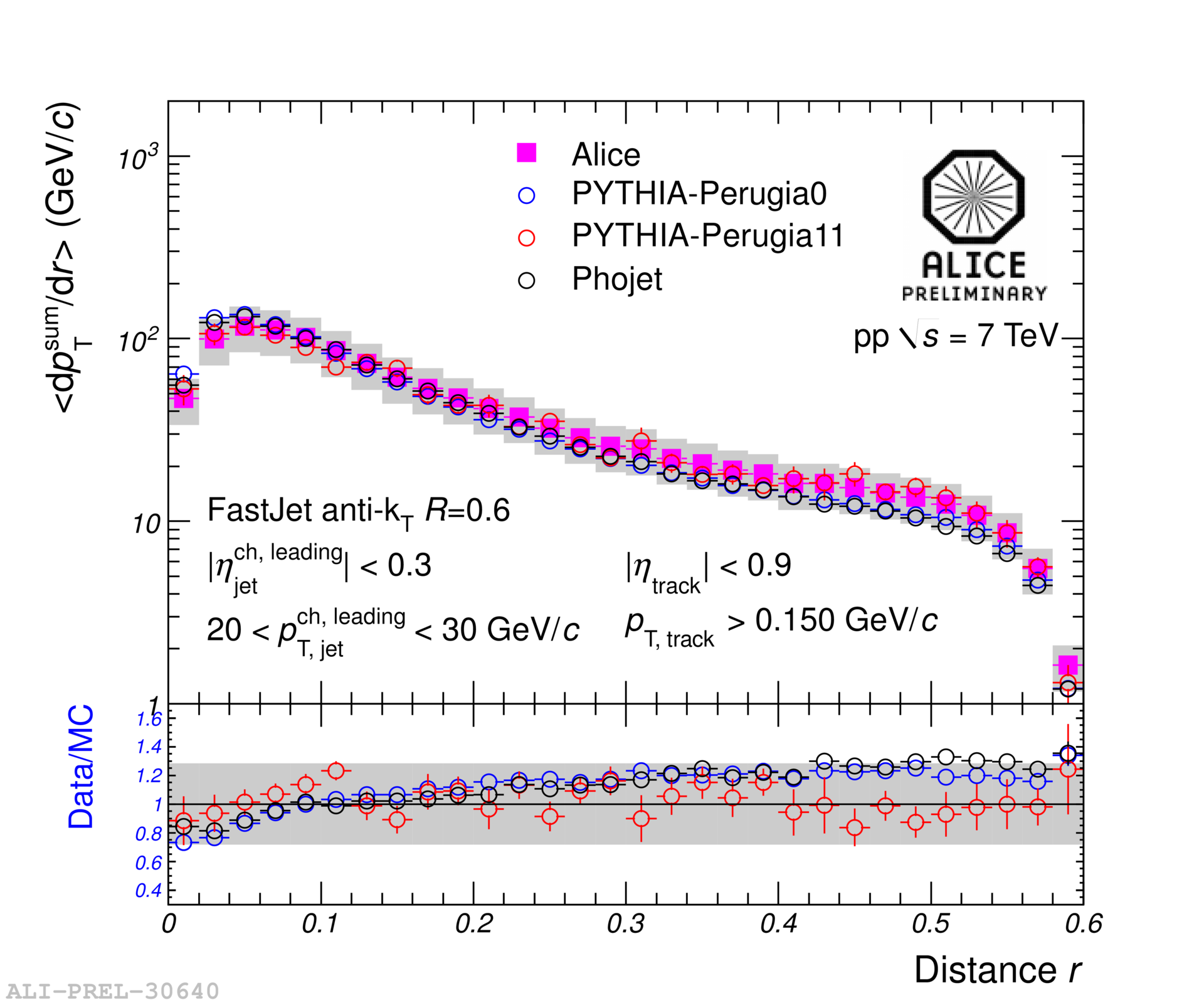}}
	\caption{ (Color online) (a) Mean charged track multiplicities of leading jet as a function of $p_{\rm{T,jet}}^{\rm{ch}}$. (b) The radial momentum distribution from a jet axis in leading charged jets in $\rm{20 \le} \textit{p}\rm{_{T,jet}^{ch}} \le 30 $~GeV/$c$ momentum bin. 
Results from pp collision at $\sqrt{s} \rm{= 7}$~TeV using the ALICE (magenta squares) and predictions from PYTHIA (Perugia0 tune) (blue circles), PYTHIA (Perugia11 tune) (red circles) and PHOJET (black circles) are shown with thes ratio of data and predictions from Monte-Carlo simulations in the bottom part. Error bars indicate statistical uncertainties and gray bands systematic uncertainties.}
	\label{res}
	\end{center}
\end{figure}

Fig.~\ref{res} shows jet shape variables. 
The charged track multiplicity of a jet, defined as a mean number of charged tracks in the leading jet, i.e. one with the largest transverse momentum in the event, belonging to given $p\rm{^{ch}_{T,jet}}$ bin, is obtained from charged tracks used for the jet reconstruction within a jet cone. 
The mean track multiplicity in such jets obtained from the analysed data is shown in Fig~\ref{res-tracks}, with respect to a charged jets' transverse momentum bin. 
Increasing value of average number of tracks is observed with increasing jet $p\rm{_T}$ bin, as previously measured by the CDF collaboration~\cite{CDF-tracks}. 
Also, predictions obtained from the PYTHIA (Perugia0 tune), PYTHIA (Perugia11 tune) and PHOJET~\cite{phojet} are shown here. 
Results from data differ by 10\% from values obtained from Monte-Carlo predictions.  

The radial momentum density is defined for jets of given $p\rm{^{ch}_{T,jet}}$ bin as a sum of momenta of charged tracks $p\rm{^{ch}_{T,track}}$ belonging to the leading jet with respect to their distance from the jet axis in $\eta - \varphi$ plane, normalized by number of jets,

\begin{equation}
\rm{
\frac{d\textit{p}^{sum}_{\rm{T}}}{d\textit{r}} \left( \textit{r} \right) = \frac{1}{2 \Delta \textit{r}} \frac{1}{N_{jet}} \sum_{jets} \textit{p}^{ch}_{T,jet} \left( \textit{r - }\Delta \textit{r, r +} \Delta \textit{r} \right),
}
\label{rad-mom-den}
\end{equation}

		where $p\rm{^{ch}_{T,jet}} \left(\textit{ r -} \Delta\textit{ r , r +} \Delta\textit{ r} \right) $ denotes the sum of $p\rm{^{ch}_{T,track}}$ of all charged tracks which are displaced from jet axis by $r \pm \Delta r $ of given jet. $N_{\rm{jet}}$ is number of leading jets in analysis.

In the Fig~\ref{res-tracks1}~
we observe a decreasing radial momentum density with decreasing distance from a jet axis. 
This suggests that most of the jet energy is concentrated around a jet axis. Comparison of the slopes of two different jet $p\rm{_T}$ bins at a small distances from jet axis indicate that high energy jets are more collimated than jets with smaller energy.
	
	\section{Conclusions}

We reported measurements of inclusive charged jet spectra in pp collisions at midrapidity,
charged track multiplicities within leading jet and the radial momentum density for jets reconstructed using anti-$k_{\rm{T}}$ algorithm. 
The reconstructed jet spectra are consistent with results obtained from the ATLAS collaboration and jet shape observables follow the trend expected from previous results of the CDF collaboration and are in reasonable agreement with predictions from PYTHIA and PHOJET calculations.

	\section*{Acknowledgements}
This work was supported by the Grant Agency of the Czech Technical University in Prague, grant No. SGS10/292/OHK4/3T/14 and by Ministry of education, youth and sports of Czech Republic, grant No. LA08015.

\end{document}